# Simultaneous Multivariate Forecast of Space Weather Indices using Deep Neural Network Ensembles

**Bernard Benson**
The University of Alabama in Huntsville
bb0008@uah.edu

**Edward Brown**
University of Cambridge
ejeb4@cam.ac.uk

**Stefano Bonasera**
University of Colorado, Boulder
stefano.bonasera@colorado.edu

**Giacomo Acciarini**
University of Strathclyde
giacomo.acciarini@gmail.com

**Jorge A. Pérez-Hernández**
Mexico National Autonomous University
jperez@icf.unam.mx

**Eric Sutton**
University of Colorado, Boulder
eric.sutton@colorado.edu

**Moriba K. Jah**
University of Texas, Austin
moriba@utexas.edu

**Christopher Bridges**
University of Surrey
c.p.bridges@surrey.ac.uk

**Meng Jin**
Lockheed Martin Solar
Astrophysics Lab
jimmeng@lmsal.com

**Atılım Güneş Baydin**
University of Oxford
gunes@robots.ox.ac.uk

## Abstract

Solar radio flux along with geomagnetic indices are important indicators of solar activity and its effects. Extreme solar events such as flares and geomagnetic storms can negatively affect the space environment including satellites in low-Earth orbit. Therefore, forecasting these space weather indices is of great importance in space operations and science. In this study, we propose a model based on long short-term memory neural networks to learn the distribution of time series data with the capability to provide a simultaneous multivariate 27-day forecast of the space weather indices using time series as well as solar image data. We show a 30–40% improvement of the root mean-square error while including solar image data with time series data compared to using time series data alone. Simple baselines such as a persistence and running average forecasts are also compared with the trained deep neural network models. We also quantify the uncertainty in our prediction using a model ensemble.

## 1 Introduction

The Sun follows an 11-year cycle, called the solar cycle, corresponding to the rise and fall in solar activity [9]. Solar radio flux proxies such as F10.7 cm, F15 cm and F30 cm, along with geomagnetic indices such as $a_p$ and $K_p$ indices are important indicators of solar variability during the solar cycle. Keeping track of solar variability is imperative due to the strong relationship between space weather and thermospheric density. Accurately forecasting space weather events can improve our estimation of thermospheric density variations, and this has a significant influence on the estimation and prediction of satellite motion. Therefore, enhancing our prediction capabilities of space weather events is pivotal for the safeguard of our space-based assets in low-Earth orbit (LEO). The F10.7 cm radio flux, in particular, is widely used in applications such as empirical modeling of the atmosphere and the ionosphere. Geomagnetic indices such as $a_p$ and $K_p$ index provide estimates of the changes in the magnetosphere and ionosphere due to solar variability. Together, the radio flux and geomagnetic indices provide information about the state of space weather. Therefore, forecasting these indices is of great importance to track space weather [4, 3].

The differential rotation of the Sun causes different latitudes to rotate at different speeds averaging to a period of approximately 27 days [20]. In this paper, we provide a simultaneous multivariate forecast of the space weather indices a full solar rotation ahead. This is expected to improve planning and estimation of quantities such as thermospheric density. Our experiments consist of two phases. First, we use time series data of the space weather indices with various forecasting methods. In this phase, a comparison study is conducted of models such as linear regression, fully-connected neural networks and long short-term memory (LSTM) [10] neural networks to find the best model that is capable of delivering accurate forecasts. Simple baselines such as a persistence model and running average forecasts are also compared with the considered models. We also quantify the uncertainty in our prediction using network ensembles. The best performing model is chosen to conduct our second phase of experiments where we include embeddings of solar image data from the NASA Solar Dynamics Observatory's (SDO) Atmospheric Imaging Assembly (AIA) instrument along with the time series data [18, 7], to investigate whether this information could enhance the accuracy of the prediction.

## 2 Related Work

Moving averages and autoregression are some of the classical techniques used for time series forecasting methods. Studies have shown that machine learning (ML) and deep learning models outperform models based on traditional approaches [1, 21]. Deep learning methods have also been applied to time series problems in heliophysics [16, 2]. For example, Topliff, Cohen and Bristow, use a single layer LSTM network to produce a multivariate forecast of geomagnetic indices with a 6-hour prediction [23]. Stevenson et al. produce a univariate forecast of F10.7 cm radio flux using an ensemble of residual networks [22]. Ensemble techniques have become increasingly popular in recent years for space weather forecasting [15]. Multi-model ensembles have been used in applications such as solar flare prediction, magnetospheric modeling, thermospheric density forecasting, and solar proxies time-series prediction [8, 14, 6, 22].

## 3 Space Weather Indices Dataset

Along F10.7 cm, solar radio flux at wavelengths such as F30 cm, F15 cm have been routinely monitored since 1951. A high degree of correlation exists between the radio fluxes and between the geomagnetic indices, which can be leveraged in a multivariate forecast. In this work, we use daily cadence time series data from https://lasp.colorado.edu/lisird/data/ for the radio flux and geomagnetic indices data from 1957 to 2021, as depicted in Fig. 1. As the average rotation period of the Sun is 27 days, our model provides forecasts one solar rotation ahead of time. For time series problems, the data needs to be pre-processed into sequences such that all the data points are formed into an input and forecast horizon pair. We use a sliding window method to sequence the data. We use four solar rotations (108 days) for the input sequence, and one solar rotation (27 days) for the forecast horizon window. For our second set of experiments we use data from the NASA Solar Dynamics Observatory [18]. A curated version of the SDO dataset, referred to as SDOML, is used to extract the images used in the experiments [7].

## 4 Model, Experiments and Results

Deep neural network ensembles can be used to quantify uncertainty as an alternative to Bayesian techniques [13]. Ensembles use the outputs of different models that can be trained in parallel to generate a robust model with the ability to quantify uncertainty in predictions [13, 5]. Lakshminarayanan et al.

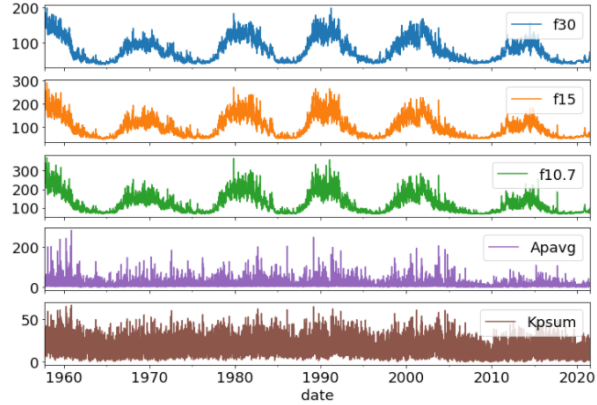

Figure 1: Solar radio flux and geomagnetic indices time series data from 1957 to 2021.

demonstrate that an ensemble of five different models improves uncertainty estimation [13]. Moreover, ensembles are generally more robust towards extreme events, as well with out-of-distribution data, coping with the overfitting of the single-point estimation models [22].

For the first phase of experiments we focus on selecting the model with the lowest root mean-square error (RMSE) and compare against baselines, including moving average and persistence models. The ML pipeline depicted in Fig. 2, visually conceptualizes model architecture, highlighting the datasets considered for the space weather indices forecasting. We use a three layer stacked LSTM with 512 features in the hidden state. The experiments are repeated using five LSTM networks with same architecture and different seeds. The data is divided such that the month of December every year is chosen as the validation subset. The Adam optimizer from PyTorch is used along with a weight decay to optimize the mean squared error loss function [11, 17]. Dropout of 20% is also used as a means of regularization to keep the model from overfitting. The ensemble networks are used to provide the forecast and the associated uncertainties.

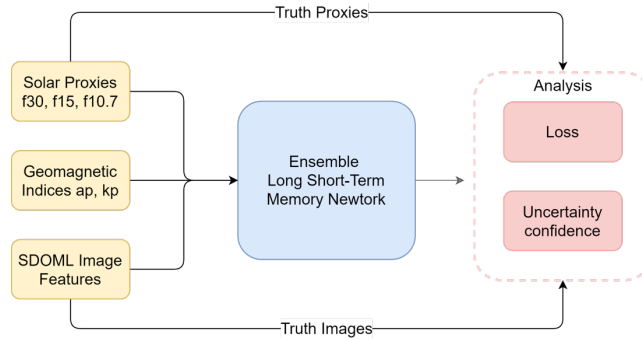

Figure 2: ML pipeline for space weather indices dataset. The inputs are passed to through the ensemble LSTM model, producing the forecast with uncertainty confidence.

For the second phase of experiments, we include image embeddings derived using a sigma variational autoencoeder from solar images as inputs along with the five features considered [12, 19]. The image data are compressed from 12 channels, each of 128x128 pixels, to a learned representation of 256 features, adding up to a total input size of 261. The input is sequenced in a similar fashion to the phase 1 experiments and passed through the ensemble of LSTM models. All experiments were run on a Google Cloud instance using a single NVIDIA Tesla T4 GPU.

The first phase experiments include testing a variety of ML architectures against simple baselines. Our results show that the stacked LSTM model performs better in terms of the RMSE metric than the other models considered. Indeed, Fig. 3a) depicts the performance of the LSTM model against other baselines and ML models. The results indicate that the LSTM model outperforms the other four models, providing the best RMSE metric for the space weather indices being forecast. The sample forecast in Fig. 3b) presents the comparison between actual and predicted data, for a sequence from the validation set. In Fig. 3b), solid lines represent the average predicted values from the ensemble LSTM model, while the dotted lines report the real measured value. The shaded regions portray the associated uncertainties in the predictions. For the solar radio flux indices the results highlight that the trained model is capable of learning the trends in data, while the associated uncertainty often bound the actual measured values, except for large spikes in data. Similar trends can be observed in the predictions for the geomagnetic indices data.

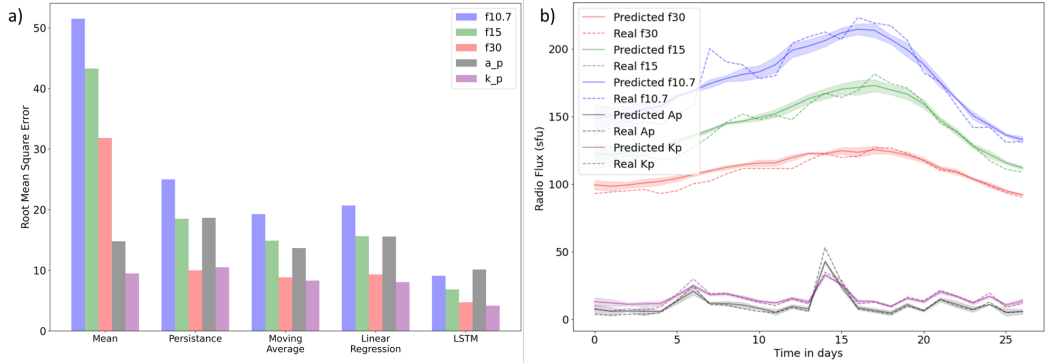

Figure 3: a) RMSE baselines for the phase I experiments. b) Sample forecast of space weather indices and associated uncertainties.

For the second phase of experiments, we add the 256 features of solar image data from the variational autoencoder to the input. The SDOML dataset provides solar images covering the period from 2010 to 2018, therefore we synchronize the dates for the space weather indices to the same duration [7]. We use these datasets as input to train a new ensemble LSTM model. Fig. 4 presents the RMSE metric for the predictions with solar features included and without solar features included, in blue and red respectively. Leveraging features from the curated SDO dataset demonstrates an improvement of 30–40% to the RMSE metric for the forecasting of solar proxies with respect to the first experiment, although reporting slightly worse results for the geomagnetic indices.

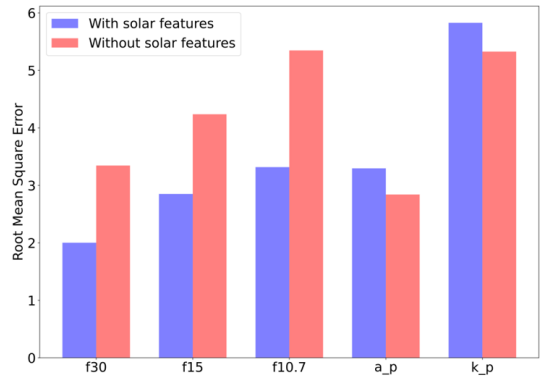

Figure 4: RMSE baselines for the phase II experiments.

## 5 Conclusions

In this study, we presented a framework for simultaneous multivariate forecasting of space weather indices using an ensemble of LSTM networks. The results of the time series work alone demonstrates the effectiveness of the model to forecast all the investigated indices while learning the trends in data. We also demonstrate the simplicity of using ensemble of LSTM models to quantify the uncertainty in the predictions. With the inclusion of the embeddings, processed with the SDOML dataset, we show the improvement in the predictions for solar indices. These initial results show the effectiveness of deep neural networks for forecasting space weather indices. We plan to continue this work by investigating different sizes of embeddings from the SDOML data and different architectures, such as transformers, to improve on our results. One limitation of this work is the limited availability of image data from the SDO. We seek to mitigate this issue by augmenting the SDO dataset with other sources of solar images.

Forecasting space weather is an important objective in heliophysics and atmospheric studies. Traditionally, space weather has been monitored via solar and geomagnetic indices, historically forecast using time-series data alone. We demonstrate the ability of deep ensemble neural networks to provide accurate forecast while quantifying the uncertainties in the predictions. The ability to provide uncertainty estimation with deep learning models, often thought of as black boxes without explainability, provides scientists and engineers with confidence in the model predictions. Additionally, by incorporating image data we demonstrate the effectiveness of using more informative data sources for space weather forecasting. Several empirical models that estimate the thermospheric density use space weather indices as inputs. As such, by providing accurate forecasts for a whole solar rotation, we can enable researchers for better planning and maneuvering of satellites, a crucial feature for collision avoidance scenarios. Eventually, improved space weather forecasts can also help in mitigating the risk of solar activity to the geo-space environment.

# 6   Acknowledgments

This work has been enabled by the Frontier Development Lab (FDL.ai). FDL is a co-operative agreement between NASA, the SETI Institute (seti.org) and Trillium Technologies Inc, in partnership with Google Cloud and Intel. We would like to thank Drs. L. Guhathakurta, M. Jin, J. van den Ijssel, E. Doornbos, A. Muñoz-Jaramillo, A. Vourlidas, I. Telezhinsky and T.S. Kelso for sharing their technical expertise and James Parr, Jodie Hughes and Belina Raffy for their support.

# References

[1] R. Adhikari and R. Agrawal. *An Introductory Study on Time series Modeling and Forecasting*. 01 2013. ISBN 978-3-659-33508-2. doi: 10.13140/2.1.2771.8084.

[2] B. Benson, W. D. Pan, A. Prasad, G. A. Gary, and Q. Hu. Forecasting solar cycle 25 using deep neural networks. *Solar Physics*, 295(5):65, May 2020. ISSN 1573-093X. doi: 10.1007/s11207-020-01634-y.

[3] C. D. Bussy-Virat, A. J. Ridley, and J. W. Getchius. Effects of uncertainties in the atmospheric density on the probability of collision between space objects. *Space Weather*, 16(5):519–537, 2018. doi: https://doi.org/10.1029/2017SW001705.

[4] N. R. Council. *Severe Space Weather Events: Understanding Societal and Economic Impacts: A Workshop Report*. The National Academies Press, Washington, DC, 2008. ISBN 978-0-309-12769-1. doi: 10.17226/12507.

[5] T. Dietterich. Ensemble methods in machine learning, 2000. In: Multiple classifier systems.

[6] S. Elvidge, H. Godinez, and M. J. Angling. Improved forecasting of thermospheric densities using multi-model ensembles. *Geosci. Model Dev.*, 9:2279–2292, 2016. doi: 10.5194/gmd-9-2279-2016.

[7] R. Galvez, D. F. Fouhey, M. Jin, A. Szenicer, A. Muñoz-Jaramillo, M. C. M. Cheung, P. J. Wright, M. G. Bobra, Y. Liu, J. Mason, and R. Thomas. A machine-learning data set prepared from the NASA solar dynamics observatory mission. *The Astrophysical Journal Supplement Series*, 242(1):7, may 2019. doi: 10.3847/1538-4365/ab1005.

[8] J. Guerra, S. Murray, and E. Doornbos. The Use of Ensembles in Space Weather Forecasting. *Space Weather*, 18(2), 2020. doi: 10.1029/2020SW002443.

[9] D. H. Hathaway. The solar cycle. *Living Reviews in Solar Physics*, 7(1):1, Dec 2010. ISSN 1614-4961. doi: 10.12942/lrsp-2010-1. URL https://doi.org/10.12942/lrsp-2010-1.

[10] S. Hochreiter and J. Schmidhuber. Long short-term memory. *Neural computation*, 9(8): 1735–1780, 1997.

[11] D. P. Kingma and J. Ba. Adam: A method for stochastic optimization, 2017.

[12] D. P. Kingma and M. Welling. Auto-encoding variational bayes. *CoRR*, abs/1312.6114, 2014.

[13] B. Lakshminarayanan, A. Pritzel, and C. Blundell. Simple and Scalable Predictive Uncertainty Estimation using Deep Ensembles, 2017. arXiv:1612.01474v3.

[14] S. Morley, D. Welling, and J. Woodroff. Perturbed Input Ensemble Modeling With the SpaceWeather Modeling Framework. *Space Weather*, 16:1330–1347, 2018. doi: 10.1029/2018SW002000.

[15] S. Murray. The Importance of Ensemble Techniques for Operational Space Weather Forecasting. *Space Weather*, 16:777–783, 2018. doi: 10.1029/2018SW001861.

[16] Z. Pala and R. Atici. Forecasting sunspot time series using deep learning methods. *Solar Physics*, 294(5), 2019. doi: 10.1007/s11207-019-1434-6.

[17] A. Paszke, S. Gross, F. Massa, A. Lerer, J. Bradbury, G. Chanan, T. Killeen, Z. Lin, N. Gimelshein, L. Antiga, A. Desmaison, A. Kopf, E. Yang, Z. DeVito, M. Raison, A. Tejani, S. Chilamkurthy, B. Steiner, L. Fang, J. Bai, and S. Chintala. Pytorch: An imperative style, high-performance deep learning library. In H. Wallach, H. Larochelle, A. Beygelzimer, F. d'Alché-Buc, E. Fox, and R. Garnett, editors, *Advances in Neural Information Processing Systems 32*, pages 8024–8035. Curran Associates, Inc., 2019. URL http://papers.neurips.cc/paper/9015-pytorch-an-imperative-style-high-performance-deep-learning-library.pdf.

[18] W. D. Pesnell, B. J. Thompson, and P. C. Chamberlin. The solar dynamics observatory (sdo). *Solar Physics*, 275(1):3–15, Jan 2012. ISSN 1573-093X. doi: 10.1007/s11207-011-9841-3.

[19] O. Rybkin, K. Daniilidis, and S. Levine. Simple and effective vae training with calibrated decoders, 06 2020.

[20] E. H. Schröter. The solar differential rotation: Present status of observations. *Solar Physics*, 100(1):141–169, Oct 1985. ISSN 1573-093X. doi: 10.1007/BF00158426.

[21] S. Siami-Namini, N. Tavakoli, and A. Siami Namin. A comparison of arima and lstm in forecasting time series. In *2018 17th IEEE International Conference on Machine Learning and Applications (ICMLA)*, pages 1394–1401, Dec 2018. doi: 10.1109/ICMLA.2018.00227.

[22] E. Stevenson, V. Rodriguez-Fernandez, E. Minisci, and D. Camacho. A deep learning approach to solar radio flux forecasting. *Acta Astronautica journal*, 2021. doi: 10.1016/j.actaastro.2021.08.004.

[23] C. Topliff, M. Cohen, and W. Bristow. Simultaneously forecasting global geomagnetic activity using recurrent networks. *CoRR*, abs/2010.06487, 2020.